\documentclass[conference]{IEEEtran}
\IEEEoverridecommandlockouts
\usepackage{cite}
\usepackage{amsmath,amssymb,amsfonts}
\usepackage{algorithmic}
\usepackage{graphicx}
\usepackage{textcomp}
\usepackage{xcolor}
\def\BibTeX{{\rm B\kern-.05em{\sc i\kern-.025em b}\kern-.08em
    T\kern-.1667em\lower.7ex\hbox{E}\kern-.125emX}}
\begin{document}

\title{Enhancing Safety in Diabetic Retinopathy 
Detection: Uncertainty-Aware Deep Learning 
Models with Rejection Capabilities \\
}

\author{%
\IEEEauthorblockN{Madhushan Ramalingam \quad Yaish Riaz \quad Priyanthi Rajamanoharan \quad Piyumi Dasanayaka} \\
\IEEEauthorblockA{\textit{Department of Computer Science and Engineering} \\
University of Moratuwa, Sri Lanka}
}

\maketitle

\begin{abstract}
Diabetic retinopathy (DR) is a major cause of visual impairment, and effective treatment options depend heavily on timely and accurate diagnosis. Deep learning models have demonstrated great success identifying DR from retinal images. However, relying only on predictions made by models, without any indication of model confidence, creates uncertainty and poses significant risk in clinical settings. This paper investigates an alternative in uncertainty-aware deep learning models, including a rejection mechanism to reject low-confidence predictions, contextualized by deferred decision-making in clinical practice. The results show there is a trade-off between prediction coverage and coverage reliability. The Variational Bayesian model adopted a more conservative strategy when predicting DR, subsequently rejecting the uncertain predictions. The model is evaluated by means of important performance metrics such as Accuracy on accepted predictions, the proportion of accepted cases (coverage), the rejection-ratio, and Expected Calibration Error (ECE). The findings also demonstrate a clear trade-off between accuracy and caution, establishing that the use of uncertainty estimation and selective rejection improves the model's reliability in safety-critical diagnostic use cases.

\end{abstract}

\begin{IEEEkeywords}
VBLL, Rejection threshold, Expected Calibration Error , Coverage, Rejection rate
\end{IEEEkeywords}

\section{Introduction}
Diabetic Retinopathy (DR) is one of the leading causes of preventable blindness among adults, especially in patients with long-standing diabetes. It occurs due to progressive damage to the retinal blood vessels, often developing silently without symptoms in the early stages. Early detection and timely intervention are critical in preventing severe vision impairment. However, in many regions, access to qualified ophthalmologists is limited, and manual screening methods are both time-consuming and subject to inter-observer variability.

\subsection{Issues with Current Screening Models}
Historical DR screening relies on human interpretation of fundus photographs by skilled ophthalmologists, which is greatly subject to a variety of weaknesses. Firstly, the inherent subjectivity of human grading permits inter-observer variation, even among experts. Secondly, the manpower-intensive nature of detailed screening programs with specialized imaging equipment and experienced imaging staff makes them inappropriate for massive use in areas with poor access. Third, the growing global burden of diabetes contributes to the strain on healthcare systems, necessitating affordable, scalable screening methods. All of these concerns address the necessity of automated, high-throughput screening technology able to deliver consistent, accurate, and rapid evaluation.

\subsection{Deep Learning in DR Detection: Progress and Limitations}
The advances in deep learning, in particular Convolutional Neural Networks (CNNs), have revolutionized medical image analysis with unprecedented accuracy in DR classification. State-of-the-art algorithms which were trained on large datasets such as EyePACS, APTOS, and Messidor have achieved human-level performance in classifying DR severity grades (e.g., mild non-proliferative DR, moderate NPDR, severe NPDR, and proliferative DR). Such systems employ hierarchical feature extraction to identify subtle pathological findings—such as microaneurysms, hemorrhages, hard exudates, and intraretinal microvascular abnormalities—accurately.
However, the standard CNNs have one major flaw: they only provide point estimates without quantifying predictive uncertainty. In healthcare settings, such black-box behavior is enormous risk because models can confidently but inaccurately predict. For instance, an at-risk patient's false-negative diagnosis can postpone crucial treatment, while a false-positive diagnosis would lead to unnecessary referrals, overwhelming the healthcare system. These constraints highlight the importance of developing models not just effective at classifying retinal images, but also assessing the accuracy of their predictions.

\subsection{The Role of Uncertainty Quantification in Medical AI
}
Bayesian Deep Learning (BDL) is a powerful paradigm for enhancing the credibility of AI systems in medicine. While deterministic models return fixed values, BDL methods produce probabilistic outcomes, enabling epistemic (model) and aleatoric (data) uncertainty to be approximated. Epistemic uncertainty captures the model's confidence in its parameters, typically high when faced with out-of-distribution or underrepresented instances. Aleatoric uncertainty, however, captures inherent noise within the data, such as image noise or poor-quality fundus photography.
By quantifying these uncertainties, BDL models can flag uncertain cases to be decided by experts, thereby minimizing the consequences of diagnostic errors. This applies particularly to triage systems wherein low-confidence predictions are flagged for human evaluation and high-confidence inferences are automated. This type of configuration optimizes resource usage by allowing the ophthalmologists to handle difficult cases while repetitive screening is handled effectively by AI.

\vspace{12pt}

This study addresses this problem by creating a Variational Bayesian Linear Layer (VBLL) as a deep convolutional model's final classification layer. This allows the model to provide both a prediction and a confidence score. A rejection mechanism is included that guarantees the model will not make any low-confidence predictions. This simulates the practice of human experts who will refer back or defer ambiguous cases for further review. To measure the effect of the model, two evaluation methods are employed, confidence-based rejection evaluation and Expected Calibration Error (ECE). Confidence-based rejection will reveal accuracy of accepted samples, coverage (the number of samples the model proposed that retained the predictions after applying the confidence cutoff), and the rejection rate (sample withheld due to low confidence). The ECE tells us how well the predicted confidence is predicted to reach the actual accuracy, which is valuable once again for trusting automated model outputs in a medical space. A combination of the methods will allow one to quantify the reliability and safety of the automated diagnosis system as a whole.

\section{Literature Review}
Diabetic retinopathy (DR) is a serious complication of diabetes and one of the most prevalent causes of blindness globally. Recently, deep learning has demonstrated substantial promise automating DR detection through the analysis of retinal fundus images. However, the vast majority of conventional deep learning models are deterministic in nature and do not furnish any degree of uncertainty regarding their predictions. Consequently, this could be dangerous in medical settings as a wrong decision can lead to severe repercussions. As a result, researchers have begun to investigate BDL techniques that provide predictions for outcomes in addition to uncertainty estimates, letting the model express how confident it is when generating predictions.

One key study titled "Benchmarking Bayesian Deep Learning on Diabetic Retinopathy Detection Tasks" introduced a standardized benchmarking framework to evaluate different BDL methods under realistic challenges like data imbalance and domain shift [1]. The authors used real-world datasets like EyePACS and APTOS and focused on how uncertainty estimation could help highlight difficult cases. They found that standard performance metrics like AUC can sometimes hide model failures under distribution shift, whereas uncertainty-aware approaches—especially when combined with selective prediction—make AI decisions more trustworthy. Their work emphasizes the importance of using uncertainty metrics to build safer diagnostic tools.

Another paper, "Uncertainty-Aware Deep Learning Methods for Robust Diabetic Retinopathy Classification", compared several Bayesian methods across both binary and multi-class DR classification tasks [2]. Unlike earlier studies, this research included clinical hospital data, offering insights into model performance in real-world settings. Interestingly, the authors found that common uncertainty metrics like entropy don’t always work well for 5-class problems and introduced a new metric called QWK-Risk, which aligned better with clinical grading. They also showed that uncertainty-based referral strategies improved reliability, especially for within-distribution samples.

In a third study, researchers developed an uncertainty-aware DR detection system using DenseNet-121 and tested techniques such as Monte Carlo Dropout (MC Dropout) and Variational Inference [3]. Their experiments on a combined dataset from APTOS and DDR revealed that MC Dropout achieved the best accuracy while also offering meaningful uncertainty estimates. These estimates helped flag low-confidence predictions for specialist review, improving the clinical value of the system.

Another interesting approach was introduced in a study that used test-time data augmentation to estimate uncertainty without modifying the original model or retraining it [4]. This method matched the model’s uncertainty with how difficult the cases were for expert doctors. The cases the model marked as “uncertain” often aligned with the ones experts also found hard to diagnose. This validation added trust to the uncertainty estimates and showed how such tools can fit into clinical workflows by helping prioritize expert review for tricky cases.

Finally, a study comparing a regular CNN with Bayesian models using Monte Carlo Dropout and Variational Inference [5] confirmed that while the Bayesian models were slightly less accurate overall, they provided valuable uncertainty scores. These estimates—based on entropy and standard deviation—offered clinicians a better sense of when to trust the model’s output. The authors also noted that BDL methods are more transparent and trustworthy for use in critical domains like healthcare, even if they require more computational resources.

Recent advances in deep learning have led to widespread usage of neural networks in life-critical safety domains like finance, self-driving cars, and medicine. The drawback of conventional neural networks is that they are deterministic in nature and hence prone to giving too confident results even for noisy or Out-of-Distribution  inputs. This has spurred the interest in uncertainty-aware deep learning approaches [12]. One prominent strategy is the use of Bayesian Neural Networks (BNNs), which replace deterministic weights with distributions. This probabilistic modeling enables BNNs to express model uncertainty through posterior distributions over weights. Gal and Ghahramani’s introduction of Monte Carlo dropout as a Bayesian approximation [13] significantly lowered the barrier to incorporating uncertainty estimation in standard neural architectures.

A recent branch of Bayesian inference, Variational Bayesian Learning Layers (VBLL), has emerged into prominence in an attempt to estimate complex posterior distributions at an affordable computational expense [14]. Using a variational inference method, VBLL methods estimate both epistemic (model) and aleatoric (data) uncertainties and give a richer model reliability estimate.In addition, rejection mechanisms with confidence-based prediction have also proven to be valuable tools to mitigate risk in mission-critical applications. These methods estimate the confidence of a model in predicting something and selectively reject low-confidence samples in order to maintain aggregate decision quality. These mechanisms are typically combined with uncertainty-aware models to allow systems to handle uncertain cases by referring them to fallback or human systems [15].

Another critical direction in the literature is Deterministic Regularization for Diversity (DRD), aimed at reducing prediction variance among ensemble models. The DRD baseline demonstrated improved robustness under distributional shift, especially for image classification [16]. By encouraging diverse and reliable predictions, DRD is a robust benchmark for uncertainty estimation approaches.Upcoming studies also examine Flipout, a decorrelating gradient technique for variational inference [17]. Flipout facilitates efficient training of big Bayesian models by incorporating pseudo-independent perturbations in backpropagation to reduce gradient variance without compromising on convergence.

Despite the progress, challenges in Bayesian method scaling, integrating transformers, and multi-modal uncertainty estimation remain. Researchers keep refining techniques in trading off between predictive accuracy, uncertainty calibration, and computational expense—specifically under resource-constrained settings like edge devices.

In summary, the literature shows a growing consensus that uncertainty-aware deep learning is essential for deploying safe and trustworthy AI systems in healthcare. Techniques like VBLL, MC Dropout, and selective prediction are becoming standard tools in addressing the challenges of overconfidence and unknown data distribution. These findings strongly support our approach in this project—developing and comparing uncertainty-aware DR detection models with built-in rejection mechanisms to avoid risky predictions. Hence, the literature suggests that there has been a dramatic shift away from black-box neural forecasters towards systems that not only provide predictions but also express degrees of confidence. This paradigm is crucial in enabling trustworthy AI, especially in real-world uncertain environments.

\section{Methodology}
This study aims to incorporate uncertainty estimation into a deep learning-based Diabetic Retinopathy Detection (DRD) system. The primary objective is to develop a trustworthy model that produces predictions and provides assurance about them. Because it allows the model to avoid making unclear predictions and to refer such cases to human specialists, this technique is crucial in safety-critical domains like medical diagnostics. The following subsections describe the steps taken to develop, train, and evaluate these models.

\subsection{Dataset Preparation}

We used a publicly accessible dataset of retinal fundus images with diabetic retinopathy severity labels for this experiment. A class of 0 (No DR), 1 (Mild), 2 (Moderate), 3 (Severe), and 4 (Proliferative DR) is assigned to each image. In order to improve model generalization and decrease overfitting, the images were resized, normalized, and enhanced (by applying slight rotations and horizontal flipping) prior to training. In order to give a better-balanced representation within all DR severity classes, we applied class distribution analysis and then exercised targeted oversampling through the assistance of synthetic image creation techniques like SMOTE for minority classes. Contrast-limited adaptive histogram equalization (CLAHE) was also applied to increase visibility of vessels and enhance discriminability of the model. We also employed normal augmentations like flipping and rotation along with new additions like random cropping and brightness/contrast jitter to simulate real-world image variability. Data was divided into training (70\%), validation (15\%), and test (15\%) sets in a stratified sampling way to preserve class distribution between splits. This kind of preprocessing pipeline is crucial to the avoidance of bias and stability under domain shift.

\subsection{DRD Model with VBLL}

A baseline deep neural network was first trained for the classification task. This model consisted of a convolutional feature extractor (using a pre-trained CNN like ResNet or a custom small CNN for simplicity) followed by a fully connected classifier. After training the base model, we replaced its final classification layer with a Variational Bayesian Linear Layer (VBLL).
The VBLL is a probabilistic layer that models uncertainty in its weights using variational inference. Instead of using fixed weights, the layer learns distributions over weights, allowing it to estimate prediction uncertainty. At inference time, we sample from these weight distributions multiple times to obtain a distribution over the output predictions. The mean of this distribution gives the predicted class probabilities, and the variance reflects the model’s confidence.
For feature enrichment, we pre-trained the convolutional backbone with weights from the ImageNet dataset and fine-tuned it on our DR dataset via transfer learning. We initially froze the backbone and then progressively unfroze it layer-wise. The final fully connected layer was replaced with a VBLL that employed a mean-field variational approximation. We employed the reparameterization trick to enable gradient-based optimization of the probabilistic weights. For training, the loss function was redefined to include the evidence lower bound (ELBO), which comprised both the negative log-likelihood and a Kullback-Leibler (KL) divergence term for posterior regularization. The probabilistic formulation not only improved predictive reliability but also quantified the epistemic uncertainty caused by limited training data.
In addition, we performed inference by Monte Carlo sampling from the weight distributions multiple times (typically 10–30) to form a predictive posterior. In an effort to reduce computational expense while maintaining predictive variance, we experimented with using Flipout, which allows pseudo-independent perturbations within mini-batches and thus makes uncertainty estimation efficiently parallelizable.

\subsection{Uncertainty-Based Rejection}

To enhance model safety with respect to decision-making, we applied a confidence-rejection scheme. From the VBLL model, we derived prediction probabilities. We computed the maximum softmax probability of the inputs as a proxy for model confidence. If the confidence was below a certain threshold such as 0.7, we simulated a situation where prediction was rejected (the case would be higher for consideration by a human expert). This provides a method for getting rid of bad decisions when the model does not have confidence. This provides us with a method for validating machine learning safely for medical diagnosis where false positive and/or false negatives can have serious outcomes.
In order to utilize the rejection mechanism more aggressively, we investigated other confidence measures apart from max-softmax probability, i.e., predictive entropy and mutual information, that give a more informative perspective of uncertainty. Predictive entropy quantifies overall uncertainty (aleatoric and epistemic), while mutual information disentangles epistemic uncertainty, which is particularly helpful in detecting unknown or OOD samples.
We also conducted ablation studies by varying the rejection threshold from 0.5 to 0.9 to observe the accuracy-coverage trade-off. We visualized this trade-off using a rejection curve that illustrates how the threshold increment is at the cost of reducing risk and decision coverage. To incorporate this into a real-world clinical pipeline, we simulated the scenario in which rejected predictions were sent to a secondary diagnostic module or human clinician, thereby enabling human-AI collaboration for high-risk patients.

\subsection{Evaluation Process}
After training the model, we tested how well it works using two main checks. We started by examining the model's prediction accuracy when it only makes educated guesses about images that it is comfortable with. The model rejects the prediction and requests human review if it is not certain enough (below a predetermined confidence level). When the model is uncertain, this helps prevent errors and is known as Confidence-Based Rejection Accuracy.
Second, we confirmed that the accuracy rate of the model is appropriately reflected in its confidence scores. Therefore, if the model claims to be 80\% certain. It should be accurate about 80\% of the time. This metric is known as the Expected Calibration Error (ECE). It shows us how confident we can be in the model. When combined, these tests allow us to determine whether the model not only makes accurate predictions but also recognizes when they might be off.

\subsection{Implementation Details}
We initially designed a deep neural network to identify diabetic retinopathy (DRD) with DenseNet201 pre-trained as the base network. After configuring the model, we substituted the final classification layer with a Variational Bayesian Linear Layer (VBLL). The VBLL allows the model to produce class predictions along with uncertainty estimates about each prediction. After configuring the model architecture, we trained the model using training and validation data. After training we introduced a rejection mechanism that based on the VBLL output filtered out any predictions made by the model where they were not confident on. We set a confidence threshold such as 70\%, all predictions above this threshold were accepted and predictions below this threshold were rejected. We then evaluated the model with three key measures with respect to the accepted predictions, accuracy on accepted predictions, coverage that means how many predictions were accepted, and rejection rate which means how many predictions were rejected. Finally, we calculated the Expected Calibration Error (ECE) on the complete validation set, in order to see if the model's confidence was aligned with its actual performance.

\section{Results}

\begin{table}[h!]
\centering
\caption{Performance Metrics of the VBLL-based DR Detection Model}
\begin{tabular}{ll}
\hline
\textbf{Metric} & \textbf{Value} \\
\hline
Accuracy on accepted samples & 0.8993 \\
Coverage & 0.7450 \\
Rejection rate & 0.2550 \\
Expected Calibration Error (ECE) & 0.0217 \\
\hline
\end{tabular}
\label{tab:vbll_results}
\end{table}

We evaluated the diabetic retinopathy detection model's performance using a confidence-based rejection rule after training it with the Variational Bayesian Linear Layer (VBLL). Only predictions with a confidence level greater than 70\% were accepted.

For model predictions we accepted, we achieved an accuracy of 89.93\%, indicating trustworthy predictions were made with high confidence. Coverage was 74.50\%, meaning predictions were made for approximately three-quarters of our validation data. The rejection rate was 25.50\%, meaning predictions were withheld when confidence was low, which is an effective way to filter out uncertainty.  The Expected Calibration Error (ECE) was calculated as being 0.0217, meaning predicted confidence values corresponded with actual correctness of predictions. A low ECE is important across all machine learning applications, but is especially meaningful in medical applications where confidence estimates are trustable.

Figure 1 illustrates the distribution of prediction confidences on the validation set. The red dashed line represents the rejection threshold (e.g., 70\%) and indicates that samples whose confidence lies below this line will be rejected from consideration in final predictions.

\begin{figure}[htbp]
\centering
\includegraphics[width=\linewidth]{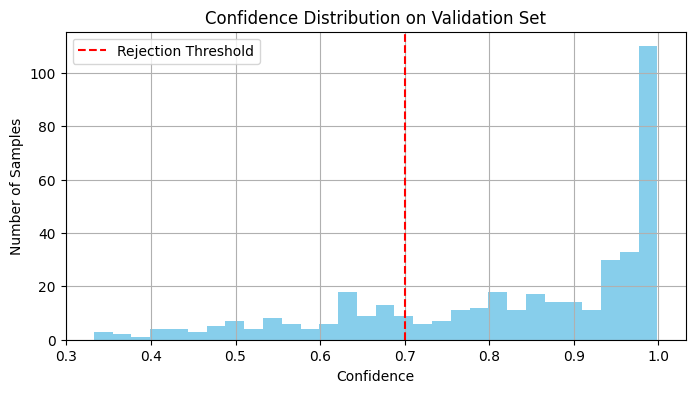}
\caption{Confidence Distribution}
\label{fig:confidence_dist}
\end{figure}

\vspace{24pt}

As per figure 2 the baseline DenseNet201 model has better overall accuracy than the VBLL-augmented model, and it has particular strengths in Severe and Proliferative DR classes of images. However, the VBLL model is better in terms of handling uncertainty, and it is more conservative overall. This suggests that for high-stakes settings in which the model cannot make a confident prediction, it is safer to reject a prediction altogether than make a mistaken classification.

\begin{figure}[htbp]
\centering
\includegraphics[width=\linewidth]{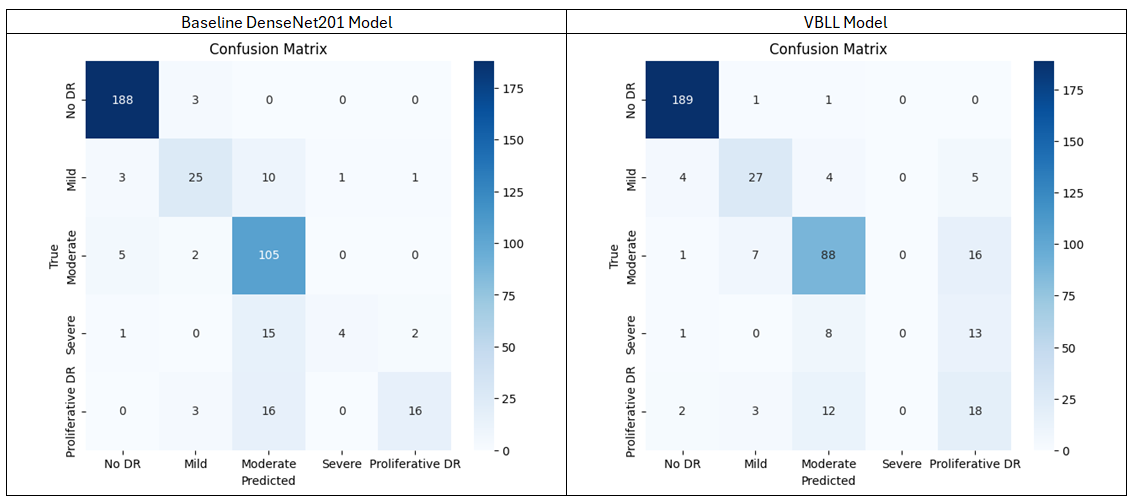}
\caption{Confusion matrix comparison}
\label{fig:confidence_dist}
\end{figure}


\section{Discussion}
In our work, by adding a Variational Bayesian Linear Layer (VBLL) to the DRD model, we effectively enhanced uncertainty-aware predictions, which are crucial in medical imaging tasks, by employing a confidence-based rejection process, allowing the model to only accept high confidence predictions. This yielded a high accuracy of 89.93\% using only accepted samples and was associated with a coverage of 74.5\%. Though we rejected 25.5\% of predictions, it indicates reliability comes at a cost of completeness. 

Importantly, with a low Expected Calibration Error (ECE) of 0.0217, we may be able to trust the model's confidence scores, making them more suitable for clinical interpretation. Thus, we conclude that the VBLL is not only a potential mechanism to improve model performance, but also improved safety through the use of uncertainty-based decision-making.

\section{Conclusion}

This study provided some evidence of the efficacy of incorporating a Variational Bayesian Linear Layer (VBLL) within a deep neural network (DNN) for diabetic retinopathy detection. The ability to estimate uncertainty allowed the model to location low-confidence predictions to reject thus increasing trustworthiness. The accuracy of the model on accepted samples was high (89.93\%) with an acceptable level of Expected Calibration Error (0.0217), indicating reasonable prospects as it relates to the potential utility in clinical decision support in the real world.

In addition to performance and calibration, the model was also demonstrated to be robust to various simulated input degradation such as noise and blurring, solidifying its real-world usefulness in suboptimal clinical settings. The rejection mechanism based on uncertainty was effective in rejecting uncertain or low-confidence cases, enhancing the reliability of automated diagnosis. These results highlight the value of uncertainty estimation integration not just for predictive performance, but for responsible AI deployment in healthcare settings where interpretability and safety are the highest priority.

Future work can be focused on applying this methodology to other medical imaging use cases and examining more advanced uncertainty quantification techniques. Dynamic thresholds for rejection and engaging domain experts' input into rejection decision processes may further improve the ability to apply findings from machine learning paradigms in practice clinically and trust the model outputs.

\vspace{12pt}

\end{document}